# Unsupervised Deep Learning for Limited-Angle STEM-EDX Tomography – Application to 3D Chemical Analysis of Phase-Change Memory Devices

Daniel del Pozo Bueno[1,*], Serge Brosset[1], Theo Monniez[1], Gabriele Navarro[1], Philippe Ciuciu[2,3], Zineb Saghi[1,*]

[1]*CEA, LETI, Univ. Grenoble Alpes, Grenoble, 38000, France.*
[2]*CEA, Neurospin, Paris-Saclay University, Gif-sur-Yvette, F-91191, France.*
[3]*Inria, MIND, Palaiseau, F-91210, France.*

*** Corresponding authors:**

Daniel del Pozo Bueno & Zineb Saghi
Commissariat à l'énergie atomique et aux énergies alternatives
MINATEC Campus
17 rue des martyrs
F-38054
Grenoble Cedex 9.
Tel: +33 (4) 38 78 48 70
Email: Daniel.delpozobueno@cea.fr
Email: zineb.saghi@cea.fr

# Abstract:

Energy Dispersive X-ray (EDX) tomography in scanning transmission electron microscopy (STEM) enables 3D compositional and elemental mapping at the nanoscale, but its use is limited by restricted tilt ranges and low-dose conditions required to avoid beam damage. Limited-angle acquisition introduces missing-wedge artefacts such as elongation and anisotropic resolution, while noisy low-dose data further degrade reconstruction quality and quantitative reliability.

Here, we introduce an unsupervised deep learning framework based on Deep Image Prior with total variation regularization (DIP-TV) for limited-angle STEM-EDX tomography. We extend it to a multi-channel formulation (DIPm-TV) that jointly reconstructs multiple elemental maps by exploiting spatial correlations. Using a synthetic three-channel phantom, we show that the method compensates for severe missing-wedge artefacts corresponding to approximately 100° of missing angular range under moderate noise, outperforming simultaneous iterative reconstruction technique and compressed sensing approaches.

We apply the method to 3D chemical analysis of Ge-Sb-Te (GST) memory devices in virgin (as-fabricated) and SET (crystalline) operational states. Samples were prepared as cross-sectional focused ion beam lamellae and acquired under a limited-angle tilt range from −40° to $+40°$ with $5°$ steps and a dose of $2.0 \times 10^5\ e^{-}\text{Å}^{-2}$. The multi-channel approach enables voxel-by-voxel elemental reconstruction using only EDX signals without external structural priors such as high-angle annular dark-field imaging. The reconstructed volumes show near-isotropic spatial resolution and reveal compositional heterogeneities associated with device operation. This approach enables 3D chemical characterization in experimentally accessible sample geometries where conventional methods fail due to severe angular limitations.

# Introduction:

Electron Tomography (ET) in Scanning Transmission Electron Microscopy (STEM) is a powerful technique for three-dimensional (3D) chemical and morphological analysis of materials at the nanoscale. When combined with Energy Dispersive X-ray spectroscopy (EDX), it enables 3D compositional and elemental mapping[1,2], which is critical for a wide range of applications, including catalysis, energy storage, semiconductors and nanoelectronics, quantum technologies, biosensing, and drug delivery[3–7].

Despite its potential, EDX tomography is constrained by the high electron dose required and the need for dense angular sampling over wide tilt ranges. In practice, tilt series are commonly acquired with coarse angular increments ($5° - 10°$) over a limited tilt range of approximately $\pm 60° - 70°$. Consequently, reconstructions suffer from streaking and Missing Wedge (MW) artefacts, including elongation along the beam direction, blurring and anisotropic resolution, resulting in reduced quantitative accuracy[8,9].

To mitigate artefacts associated with sparse-view and limited-angle conditions, multimodal (MM) reconstruction strategies have been proposed that combine STEM-EDX and STEM-High-Angle-Annular Dark-Field (HAADF) signals, exploiting the higher signal-to-noise ratio (SNR) and spatial resolution of STEM-HAADF imaging as structural priors to improve STEM-EDX reconstructions.

Following this idea, bimodal and regularized reconstruction frameworks have been proposed[10–12] and fused multimodal electron tomography (fused MM-ET)[13–15] has emerged as a state-of-the-art strategy combining EDX, Electron Energy Loss Spectroscopy (EELS) and HAADF signals. MM-ET improves dose efficiency by densely sampling the HAADF tilt series (typically $< 5°$ increments), while spectroscopic data are acquired at coarser angular steps (e.g., $10° - 15°$). Reconstruction quality is improved through HAADF-derived structural constraints that compensate for sparse and noisy spectroscopic data.

More recently, Deep Learning (DL) approaches have been explored for low-dose spectroscopic STEM tomography. Cha et al.[16] introduced an unsupervised cycleGAN-based denoising framework combining EDX and HAADF tilt series acquired from $-60°$ to $+60°$ with $10°$ increments, and ProjectionGAN-based method for reducing MW artefacts using symmetry constraints in the projection domain.

In this work, we address two key challenges: (1) insufficient HAADF Z-contrast for effective multimodal reconstruction, and (2) severely limited tilt ranges arising from sample geometry (e.g., planar samples or those exhibiting strong structural overlap), further exacerbated in in-situ experiments by hardware constraints (e.g., dedicated holders, liquid cells, or heating chips) that restrict the accessible angular range[17,18].

To overcome these limitations, we introduce an unsupervised DL framework based on the Deep Image Prior (DIP)[19–22] for highly limited-angle, low-dose STEM-EDX tomography requiring no external training data, HAADF priors, or symmetry constraints. Building on the DIP-TV framework proposed in[23], we extend it to a multi-channel formulation (DIPm-TV) that exploits inter-element correlations to improve robustness under low SNR conditions. DIPm-TV acts as an implicit regularizer by optimizing an untrained convolutional neural network (CNN) directly on measured EDX projections through the forward model (i.e., Radon transformation), enabling simultaneous denoising and artefact suppression. We demonstrate robust reconstruction performance for tilt ranges as limited as $\pm 40°$ (MW$\sim$100°), with simulations indicating stability up to $\sim$140° MW under moderate noise conditions.

As a proof of concept, we apply DIPm-TV to EDX tomography of semiconductor devices, where nanoscale composition and elemental distribution critically determine device performance[6,24]. Traditionally, such systems have been studied using needle-shaped specimens to enable full angular sampling ($\pm$90°), thereby eliminating the MW. However, as device dimensions shrink and architectures become more complex, this preparation strategy becomes increasingly impractical and often incompatible with realistic device geometries. In contrast, Focused Ion Beam (FIB)-prepared lamellae provide a more versatile and widely applicable approach, preserving the device environment but at the cost of severely restricted tilt ranges due to geometric constraints and projection overlap at high tilt angles.

Here, we focus on phase-change memory (PCM) devices based on Ge-Sb-Te (GST) alloys, which are promising candidates for next-generation non-volatile memory technologies[25,26]. Their performance is governed by nanoscale features such as elemental migration and grain composition evolution across different operational states. Access to this information in 3D is essential, as it cannot be reliably inferred from two-dimensional (2D) projections due to thickness integration and overlapping structures along the beam direction. Applied to a comparative study of PCM devices in virgin and SET states, DIPm-TV enables voxel-by-voxel, semi-quantitative 3D chemical mapping with near-isotropic spatial resolution, extending EDX tomography to semiconductor STEM specimen geometries beyond the reach of conventional approaches.

# Results & Discussion:

## DIPm-TV validation on a synthetic phantom:

To quantitatively evaluate DIPm-TV, we designed a 3D synthetic phantom consisting of a matrix with two embedded structures of different intensities (Figure S1). Tilt series were generated using the ASTRA Toolbox, and Poisson noise (SNR ≈ 20 dB) was added to emulate realistic experimental conditions. A high-quality SIRT reconstruction derived from a [−90°:1°:+90°] tilt series was taken as the reference for benchmarking SIRT, CS-TV and DIPm-TV under various limited-angle conditions.

Figure 1 evaluates reconstruction quality as a function of tilt range (180° to 40°) for the synthetic phantom, using the structural similarity index measure (SSIM) and normalized root mean square error (NRMSE) as metrics, together with visual assessment of representative cross-sections. Metrics are averaged across the three signals, with uncertainty bands reflecting geometry-dependent variability. While absolute values are partially influenced by background regions, the spread of the uncertainty bands provides a measure of robustness to structural variations. SIRT and CS-TV reconstructions deteriorate significantly with decreasing tilt range and exhibit increased variability, consistent with their sensitivity to MW artefacts. In contrast, DIPm-TV consistently achieves the highest SSIM and lowest NRMSE across all tilt ranges, with narrow uncertainty bands indicating robust, geometry-independent performance.

Visual inspection of CS-TV and DIPm-TV reconstructions complements these findings (Figure 1e-t and Figure S2). CS-TV reconstructions progressively degrade with decreasing tilt range, showing elongation artefacts and blurring of both the matrix and internal features. By contrast, DIPm-TV preserves sharp interfaces and the cuboid matrix morphology even at $40°$ tilt range (SSIM $= 0.95$ vs $0.72$ for CS-TV). Similar behaviour is observed for the spherical and ellipsoidal inclusions, which remain structurally consistent under DIPm-TV across all tilt ranges. Figure S3 shows line profiles in the MW direction through the reference and the SIRT, CS-TV and DIPm-TV reconstructions for the [-40°:5°:+40°] scenario. Beyond structural improvements, DIPm-TV also demonstrates superior intensity recovery accuracy compared to SIRT and CS-TV – a critical requirement for quantitative AET.

These results demonstrate that DIPm-TV effectively mitigates MW artefacts and delivers robust, geometry-independent performance by leveraging multi-channel correlations, making it particularly well suited for limited-angle STEM-EDX tomography of FIB-prepared lamellae.

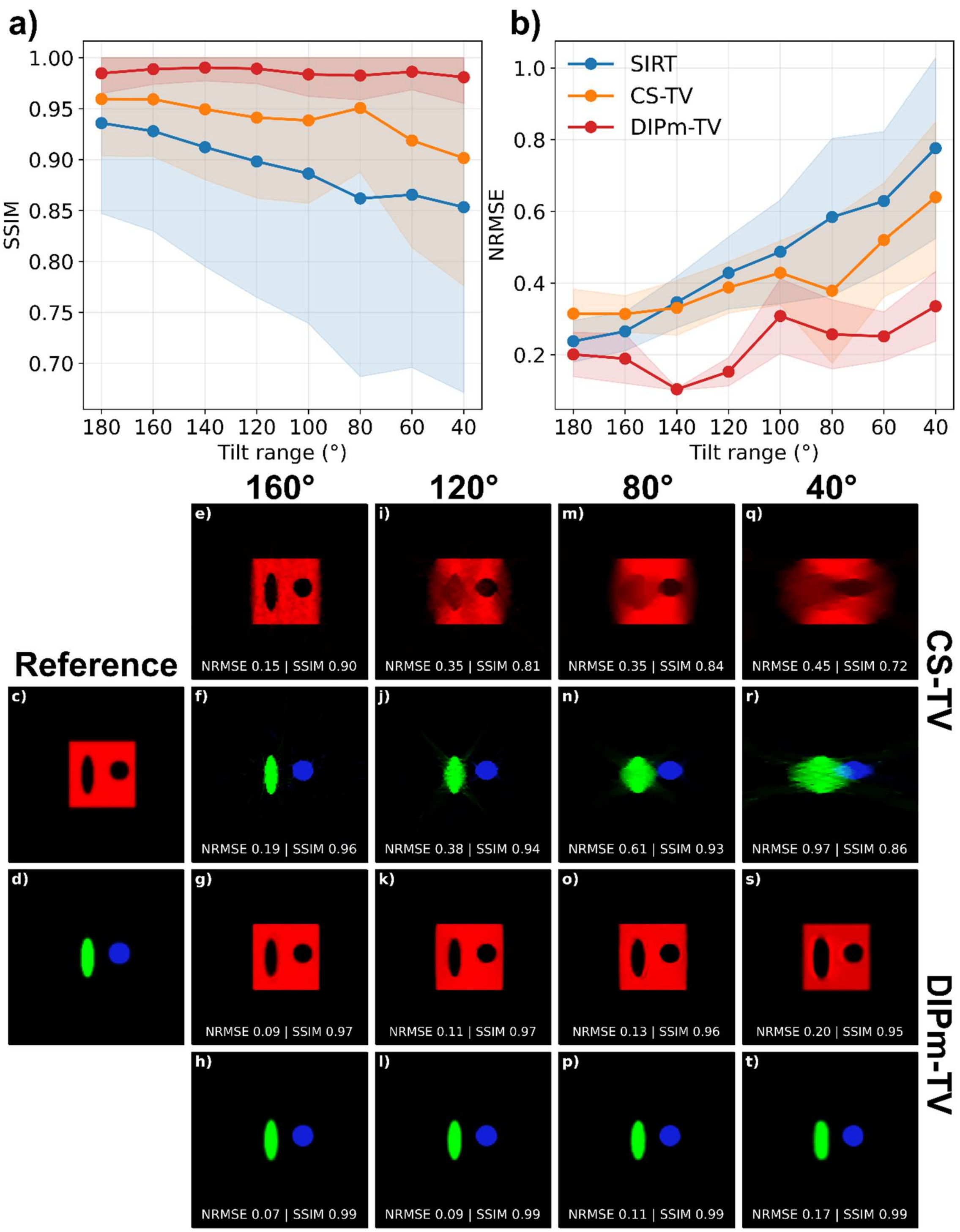

Figure 1. Effect of limited tilt range on reconstructions of a simulated phantom. Evolution of volumetric metrics: (a) SSIM and (b) NRMSE as a function of the tilt range for SIRT, CS-TV and DIPm-TV. (c-d) Reference slices of the three-channel phantom, showing the matrix and embedded inclusions. (e-t) Representative cross-section slices for different tilt ranges: upper rows correspond to CS-TV reconstructions, and lower rows to DIPm-TV reconstructions. The MW is displayed horizontally.

## DIPm-TV for limited-angle STEM-EDX tomography of PCM memory devices:

In this section, DIPm-TV was applied to perform 3D chemical mapping of two PCM devices in the virgin and SET states. Samples were prepared as conventional FIB lamellae and mounted on a Fischione single-axis tomography holder such that the tilt axis was aligned parallel to the GST layer. STEM-EDX and STEM-HAADF tilt series were acquired from $-40°$ to $+40°$ in $5°$ increments (17 projections; one was discarded due to acquisition corruption, yielding 16 projections for reconstruction). The accessible tilt range was constrained by overlap from surrounding device structures.

HAADF-STEM images acquired at $0°$ (Figure 2a, f) show a $\sim 50$ nm-thick GST region, consistent across both devices. STEM-EDX data were acquired using a multiframe strategy, corresponding to a total dose of $2 \times 10^5\, e^- \,\text{Å}^{-2}$. The resulting chemical maps exhibited SNR values of $21.6\ \pm\ 4.1$ dB and $20.0\ \pm\ 4.2$ dB for the virgin and SET devices, respectively, with element-dependent variations. Detailed per-element SNR values are provided in Supplementary Table S1.

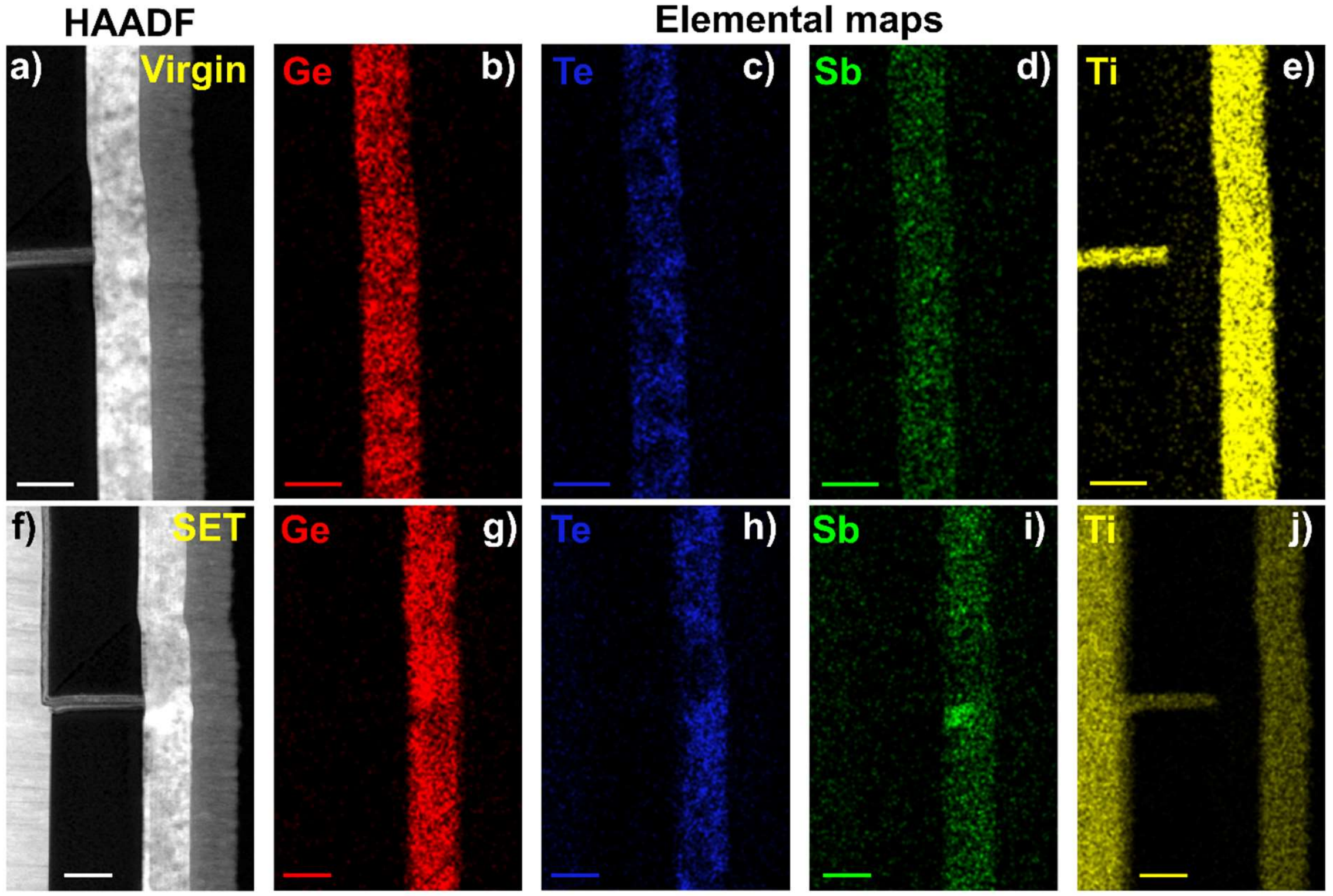

Figure 2. HAADF-STEM image and corresponding EDX elemental maps (Ge, Te, Sb, Ti) acquired simultaneously at 0° for the two PCM devices. Scale bars: 50 nm. The tilt axis is vertical (parallel to the GST layer).

Figure 3 shows the central XZ, YZ, and XY cross-sections of the reconstructed chemical volumes obtained using CS-TV and DIPm-TV, where Z corresponds to the MW direction (that is, the electron beam direction). MW artefacts are prominent in the CS-TV reconstructions, appearing as a horizontal elongation in XZ slices (Figure 3a, g) and vertical elongation in YZ slices (Figure 3c, i). In addition, the CS-TV reconstructions exhibit pronounced streaking artefacts arising from the large tilt increment, and low SNR despite TV regularization.

In contrast, DIPm-TV largely suppresses these artefacts and preserves a compact morphology with sharp interfaces. This is particularly apparent in the XZ cross-sections (Figure 3b, h), where DIPm-TV recovers a well-delimited square morphology consistent with the expected device morphology[27]. In the YZ plane (Figure 3d, j), DIPm-TV restores a more faithful structure with reduced elongation and improved interface definition. In the XY plane, which is unaffected by MW artefacts, DIPm-TV further enhances lateral resolution and reduces inter-grain mixing compared to CS-TV, highlighting its intrinsic denoising capability. SIRT reconstructions, exhibiting even more pronounced artefacts than in CS-TV, are shown in Figure S4.

Moreover, the spatial resolution achieved by the different reconstruction methodologies was evaluated using power spectral density (PSD) cut-off analysis[13] (Table S2). For SIRT and CS-TV, the in-plane (XY) resolution ($5.3 - 5.8\,nm$) remains close to the sampling limit defined by the Nyquist criterion ($4.3\,nm$), indicating that lateral resolution is primarily acquisition-limited. By contrast, the MW direction exhibits substantially degraded resolution ($7.0 - 7.8\,nm$), corresponding to anisotropy ratios of $1.29 - 1.35$, consistent with the theoretical elongation factor imposed by the limited tilt range, $E_l(\theta_{max}) = 1/\cos(\theta_{max}) \approx 1.3$.

DIPm-TV achieves comparable lateral resolution ($5.2 - 5.5\,nm$) while substantially improving the resolution along the Z direction ($5.7 - 5.9\,nm$), yielding near-isotropic reconstructions with anisotropy ratios of $1.06 - 1.08$. A complete summary of the extracted resolution values is provided in Table S2, with the corresponding PSD profiles shown in Figures S5-S7.

Notably, this improvement is consistently observed across all elemental channels despite pronounced differences in SNR. This demonstrates that the multichannel formulation effectively balances information across signals, enabling robust reconstruction even for low-count elements.

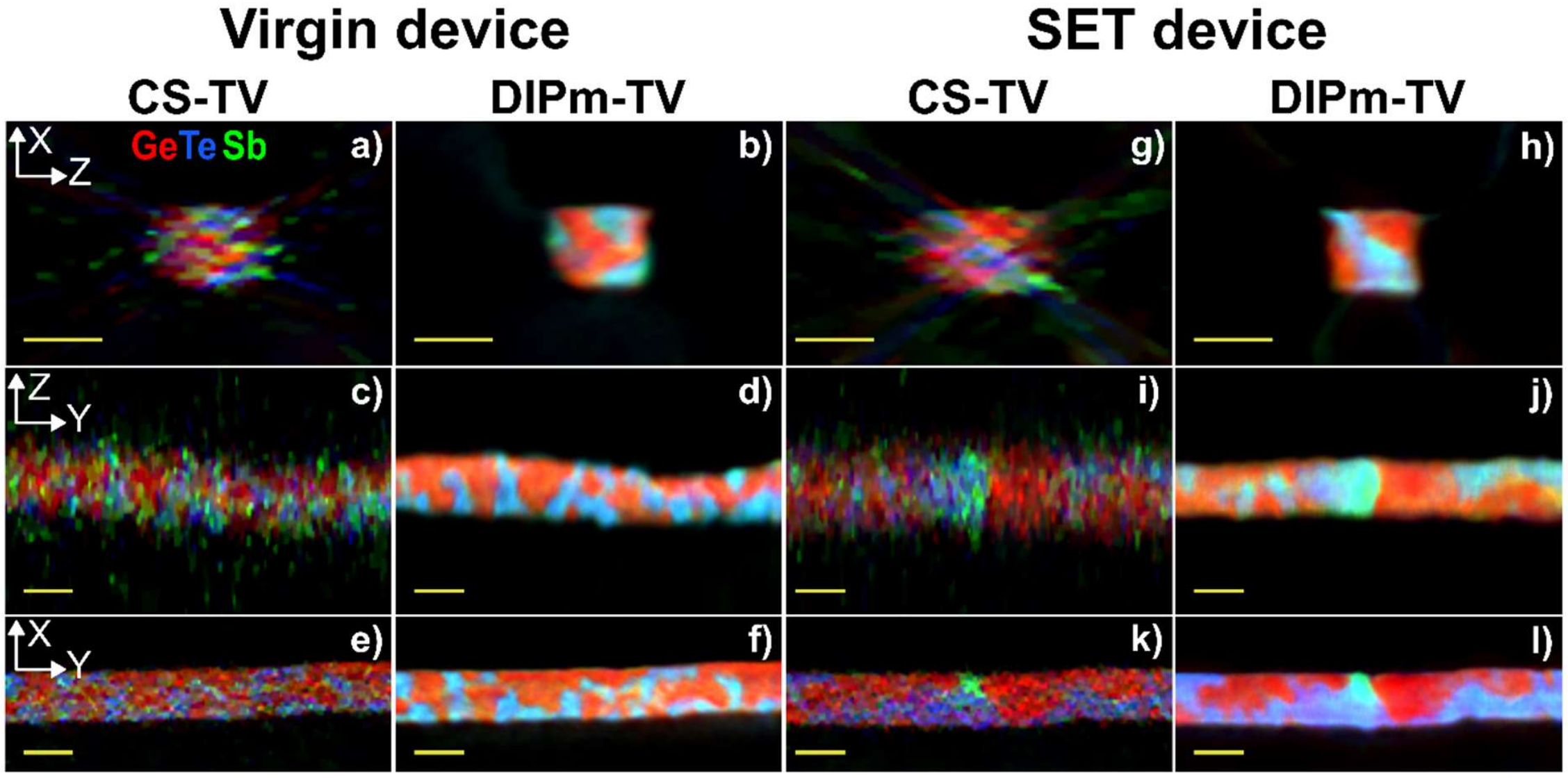

Figure 3. Central XZ, YZ, and XY cross-sections of the 3D elemental volumes obtained using CS-TV, and DIPm-TV for the virgin (left) and SET (right) devices. Scale bars: 50 nm. The MW is in the Z direction.

Volume renderings of the DIPm-TV reconstructions are shown in Figure 4a and 4d, which reveal the 3D elemental distribution and the spatial relationship with respect to the TiN heater structure (visualized in grey). The virgin device exhibits a finer, more homogeneous granular morphology (Figure 4a), typical of an as-fabricated device, whereas the SET-state device shows a coarser and more spatially organized structure (Figure 4d), with more pronounced Ge and Sb/Te segregation and the emergence of a localized Sb-rich region near the TiN heater. This structural evolution is consistent with Ge diffusion and chemical phase reorganization driven by coupled electrothermal effects during SET programming[28,29].

To further quantify these compositional differences, atomic percent concentrations (at.%) of the elements of interest were estimated using the Cliff-Lorimer (CL) quantification method[30]. Ternary diagrams and K-means clustering analysis were performed on the quantified volumes (Figure 4b and 4e). For comparison, ternary diagrams derived from SIRT and CS-TV reconstructions are shown in Figure S8. In the virgin device (Figure 4c), two main compositional domains are identified: a Ge-rich phase (about 59 vol.%) and a GST-like phase (about 41 vol.%). In the SET device (Figure 4f), two similar phases are found and an additional Sb-rich cluster (about 2 vol.%) emerges, spatially localized near the TiN heater. While the Ge-rich phase remains compositionally stable, the GST-like phase exhibits a slight Ge enrichment accompanied by Te reduction, indicating subtle redistribution during programming. In both cases, the compositional distributions follow an elongated trend along the Ge-Te axis in the ternary space, consistent with the known behaviour of the studied materials.

These results demonstrate that DIPm-TV enables 3D semi-quantitative chemical mapping, capturing both morphological and compositional changes associated with device operation, including subtle elemental redistribution and the emergence of localized compositional domains.

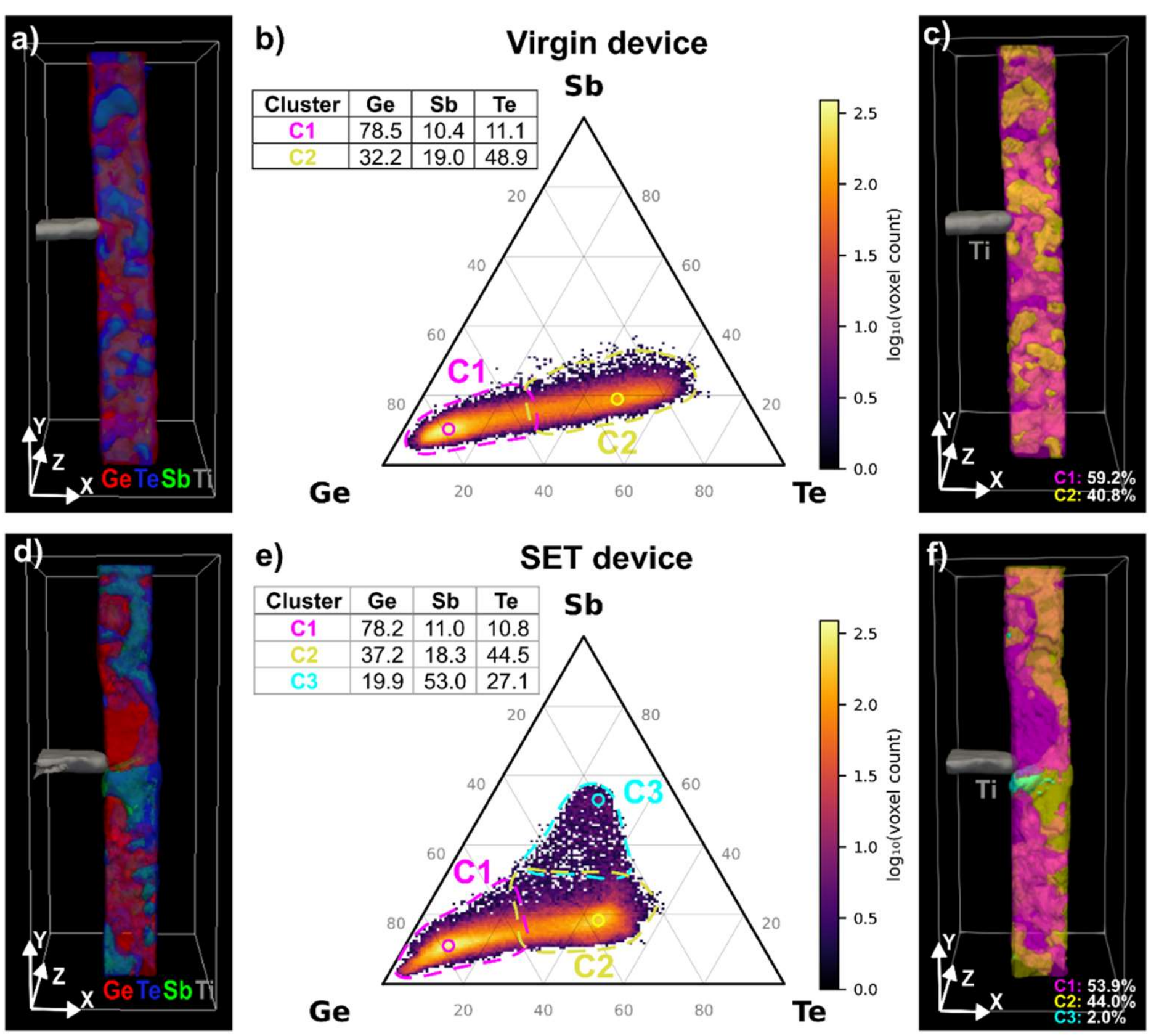

Figure 4. DIPm-TV reconstruction renderings with elemental contributions color-coded −Ge (red), Te (blue), Sb (green), and Ti (grey)− of the virgin (a) and SET (d) devices. (b, e) Ternary diagrams showing voxel-wise compositional distributions ($\log_{10}$ voxel count colour scale), with K-means clusters highlighted by dashed lines. The representative compositions shown correspond to the maximum voxel-density regions, as reported in the inset tables. (c, f) 3D spatial cluster maps of the reconstructed volumes, with cluster volume fractions indicated. C1 (magenta): Ge-rich phase; C2 (yellow): GST-like phase; C3 (cyan): Sb-rich phase.

## Discussion:

The results presented in this work demonstrate the potential of DIPm-TV for quantitative 3D characterization of semiconductor devices prepared as FIB lamellae. In particular, this approach enables the reconstruction of PCM devices under severely restricted tilt ranges, while mitigating MW artefacts and maintaining robust performance under low-dose conditions. Importantly, these findings highlight the added value of voxel-wise analysis: in contrast to conventional 2D EDX projections, where grain superposition prevents local chemistry, 3D reconstructions provide direct access to nanoscale compositional heterogeneity.

Compared with state-of-the-art approaches, DIPm-TV operates in a similar dose regime ($\approx 2 \times 10^5\ e^-/\text{Å}^2$) to that reported for multimodal strategies ($\approx 10^4$– $10^6\ e^-/\text{Å}^2$)[13,14,16]. However, most existing methods were successful for MW values below $\lesssim 60°$, whereas DIPm-TV demonstrates robust performance under MW conditions approaching $100°$, without requiring external structural priors such as HAADF images. This substantially extends the applicability of EDX tomography to samples where such priors are unavailable. Notably, the full EDX tilt series (16 projections over a $\pm 40°$) was acquired in $\sim 74$ minutes ($\sim 2$ hours including alignment), highlighting the practicality of the approach for routine acquisition.

The achieved spatial resolution ($5 - 5.8$ nm) approaches the Nyquist limit (4.3 nm) and is nearly isotropic, despite relying exclusively on EDX signals. This improvement reflects both enhanced structural recovery and the suppression of noise-induced high-frequency components, consistent with the implicit regularization inherent to DIP-based frameworks.

These results also point to clear avenues for further improvement, particularly in low-SNR regimes. Incorporating Poisson-aware data fidelity terms beyond the current L2 formulation, together with emerging DL approaches such as implicit neural representations (INRs) and plug-and-play (PnP) priors, may further improve reconstruction quality in low-dose experiments. Although preliminary INR-based tomographic reconstructions have been reported[31], their robustness across realistic spectroscopic datasets remains to be established. More generally, acquisition-aware frameworks that explicitly incorporate counting statistics and detector response across both spectral and reconstruction stages represent a promising direction for future work.

The performance of DIPm-TV depends on several parameters, including network architecture, learning rate, number of iterations, and TV regularization weight ($\lambda_{TV}$). Overfitting may occur after an initial signal-learning phase[19], and no universally accepted stopping criterion currently exists[32]. While this remains an open problem, in this work we adopt an empirical strategy to address it. The optimal $\lambda_{TV}$ value is strongly dependent on the noise level: higher noise requires stronger regularization, whereas lower-noise datasets benefit from reduced regularization to preserve fine structural details. Optimal performance was obtained with very small values ($\lambda_{TV} \approx 10^{-9} - 10^{-10}$). As a practical guideline, $\lambda_{TV}$ was chosen such that the TV term acted as a weak regularizer relative to the data-fidelity term after convergence, ensuring that the reconstruction remains predominantly data-driven. For the stopping criterion, intermediate reconstructions were systematically stored and evaluated, and the optimal solution was selected within the convergence plateau of the loss function, balancing under- and overfitting[20].

Additional limitations arise from the spectral processing itself. Elemental intensities are extracted using a commercial software package, which limits transparency in spectral processing and constrains rigorous error propagation. This issue becomes more critical at low dose, where reduced photon counts increase variance and amplify sensitivity to noise. These considerations motivate the use of more transparent spectral approaches, such as Bayesian or multiscale methods that explicitly account for Poisson statistics[33].

Furthermore, quantification based on the CL method introduces intrinsic uncertainties ($\sim 10{-}\,15\ at.\%$), arising from k-factor accuracy and absorption effects, particularly for low-energy X-rays[30,34,35]. As a result, the reported compositions should be interpreted in a semi-quantitative manner, with emphasis on relative trends rather than absolute values. No voxel-wise absorption correction was applied in this study, justified by the relatively high energy lines selected and the moderate lamella thickness. Nevertheless, iterative 3D absorption correction strategies, such as those proposed by Burdet et al.[36], represent a natural extension. In this context, the improved structural fidelity of DIPm-TV reconstructions compared with SIRT or CS-based methods is expected to provide a more accurate initial estimate, enabling more reliable absorption correction.

From a broader perspective, these results demonstrate that spectroscopic ET can be extended to realistic samples geometries, enabling direct 3D chemical characterization under experimentally accessible conditions. The method is also particularly relevant for in-situ ET, where experimental constraints further restrict the accessible tilt range[17].

Finally, although this study focuses on STEM-EDX tomography, the proposed framework is directly applicable to EELS tomography and combined EDX-EELS tomography, broadening its applicability across spectroscopic modalities.

# Conclusions:

In this work, we have demonstrated that DIP-TV in a multi-channel formulation provides a robust and versatile framework for 3D reconstruction in EDX tomography under severely limited-angle conditions. The method effectively mitigates MW artefacts without external structural priors, thereby extending EDX tomography to FIB-prepared lamellae of functional materials and semiconductor devices. The multi-channel formulation jointly reconstructs elemental signals within a shared network architecture, exploiting inter-channel spatial correlations. This enables low-SNR channels to benefit from structural information present in higher-SNR channels, yielding consistent reconstruction quality across elements despite large differences in signal intensity. The implicit regularization imposed by the convolutional architecture further acts as an effective denoising prior, suppressing noise-induced artefacts without explicit noise modelling.

Beyond methodological performance, DIPm-TV enables voxel-by-voxel, semi-quantitative reconstruction of elemental distributions with near-isotropic spatial resolution under realistic low-dose conditions. Applied to GST-based PCM devices in the virgin and SET states, it reveals nanoscale compositional variations associated with device operation that are inaccessible to conventional reconstruction methods. In particular, a spatially localized Sb-rich phase is observed near the TiN heater in the SET device, consistent with electrothermal-driven elemental redistribution during programming. This highlights the importance of reliable 3D compositional reconstruction in nanoscale semiconductor devices.

These capabilities provide a route towards quantitative interpretation of structure-composition relationships in complex semiconductor systems and establish a foundation for advanced 3D chemical imaging in next-generation devices.

# Methods:

## Sample synthesis and preparation:

In this work, two GST-based PCM devices were investigated. The devices were fabricated using a multilayer deposition strategy for chalcogenide alloys, enabling precise control over composition, thickness, and structural stability during growth[27]. This approach is particularly suitable for GST alloys, which are promising candidates for embedded automotive memory applications due to their improved thermal stability and reduced variability[27,37]. The two devices correspond to distinct operational states: a virgin (non-programmed) device and a programmed device in the SET state (i.e., crystalline phase). Cross-sectional specimens were prepared by FIB milling using standard lift-out procedure. Final thinning was performed at low accelerating voltage to minimize ion-beam-induced damage and amorphization.

## Experimental acquisition and pre-processing:

STEM-EDX tomography experiments were performed using a Thermo Fisher Scientific Titan Themis microscope operated at 200 kV, equipped with a probe corrector and a Super-X EDX system composed of four silicon drift detectors. Tilt series were acquired over a limited angular range of $\pm 40°$ with 5° increments (17 projections), using identical acquisition conditions for both devices. One projection was discarded in both cases due to acquisition corruption, yielding 16 projections for reconstruction.

At each tilt angle, STEM-EDX spectrum images were acquired simultaneously with HAADF-STEM images using Velox software (Thermo Fisher Scientific). Elemental maps were extracted by integrating the net counts corresponding to the Ge Kα, Te Lα, Sb Lα and Ti Kα emission lines. A summary of the acquisition parameters and corresponding electron dose is provided in Table 1.

Projection alignment was performed in Fiji using the MultiStackReg package with translation-only registration[38]. Transformation parameters obtained from the Ge maps were subsequently applied to the remaining elemental channels to preserve a consistent projection geometry across the dataset. Following alignment, the elemental maps were spatially cropped and resampled to 176 × 112 pixels (binning factor of 2) and normalized independently by channel using their maximum intensity prior to reconstruction.

| STEM-EDX settings | |
|---|---|
| **Accelerating voltage** | 200 kV |
| **Beam current** | 115 pA |
| **Camera length** | 110 mm |
| **Tilt range** | $-40°$ to $+40°$ |
| **Tilt increment** | $5°$ |
| **Frame size (before binning)** | $400 \times 300$ |
| **Pixel size (before binning)** | 10.67 Å |
| **Dwell time** | 40 $\mu s$ |
| **Acquisition type** | Multi-frame (50 frames) |
| **Time per frame** | 5.57 $s$ |
| **Cumulative time per tilt angle** | 4 $min$ 38.5 $s$ (50 frames) |
| **Fluence rate** | $6.30 \times 10^6\ e^{-}Å^{-2}s^{-1}$ |
| **Fluence per frame** | $2.52 \times 10^2\ e^{-}Å^{-2}$ |
| **Cumulative fluence (50 frames)** | $1.26 \times 10^4\ e^{-}Å^{-2}$ |
| **Total fluence (16 projections)** | $2.0 \times 10^5\ e^{-}Å^{-2}$ |

Table 1. Experimental settings for STEM-EDX tomography acquisitions, and the corresponding electron fluence (dose).

## Simulated data generation and quantitative analysis:

The phantom consisted of a matrix and two embedded structures with distinct morphologies and intensities, reproducing characteristic geometrical features of the experimental system. Poisson noise was subsequently added to emulate experimental acquisition conditions. Next, ASTRA Toolbox[39,40] was used to compute projection data and simulate tilt series under different acquisition scenarios. Limited-angle conditions were emulated by varying the tilt range from $\pm 20°$ to $\pm 90°$ in $10°$ steps, while maintaining a constant tilt increment of $5°$.

Quantitative analysis and image quality metrics were then evaluated, using as reference the SIRT reconstruction obtained from a fully sampled tilt series spanning $-90°$ to $+90°$ with $1°$ angular increments. SSIM and NRMSE were computed using the scikit-image implementation[41]. SNR was estimated from signal and background regions as described in the Supplementary Information.

## DIPm-TV reconstruction framework and training:

DIP-TV formulates the reconstruction as a regularized inverse problem in which the 3D volume is parameterized by an untrained CNN optimized directly against the measured projections[19–23], without requiring training data or external structural priors. The reconstruction emerges from the implicit regularization imposed by the convolutional architecture itself, combined with explicit TV regularization[21–23]. The reconstructed volume is obtained by optimizing the CNN parameters through the forward projection model:

$$\hat{\theta} = argmin_{\theta} \left\{ \frac{1}{2} \left|\left| PF_{\theta}(z) - y \right|\right|_2^2 + \lambda \cdot \left|\left| \nabla F_{\theta}(z) \right|\right|_1 \right\}, (1)$$

where $y$ denotes the measured projections, $P$ the projection operator, $F_{\theta}(z)$ the CNN output parameterized by network weights $\theta$ and fixed random input $z$, and $\lambda$ the TV regularization weight.

Building upon previous DIP-TV formulations[19,23], we extend the approach to a multi-channel reconstruction strategy (DIPm-TV), in which all elemental signals are reconstructed jointly. This formulation exploits spatial correlations between elemental distributions, allowing low-SNR channels to benefit from structural information present in higher-SNR channels. The optimization problem is then expressed as:

$$\hat{\theta} = \underset{\theta}{\operatorname{argmin}} \sum_{c=1}^{N} \left\{ \frac{1}{2} \left|\left| PF_{\theta}^{c}(z) - y^{c} \right|\right|_2^2 + \lambda \cdot \left|\left| \nabla F_{\theta}^{c}(z) \right|\right|_1 \right\}, (2)$$

where $y^c$ and $F_{\theta}^{c}(z)$ denote the projections and reconstructed volume for channel $c \in \{1, \cdots, N\}$, respectively, with $N$ representing the total number of channels. Moreover, inter-channel correlations are exploited through the shared network weights $\theta$, enabling the joint reconstruction of all elemental volumes within a single optimization. The reconstructed multi-channel volume is defined as:

$$\hat{x} = \left[ F_{\hat{\theta}}^{1}(z), \cdots, F_{\hat{\theta}}^{N}(z) \right]. (3)$$

The reconstruction framework was implemented in PyTorch using a 3D U-Net-like architecture. Forward projections were computed using Tomosipo[42]. Optimization was performed using AdamW[43]. CS-TV reconstructions were obtained using pysap-etomo[44]. All reconstructions were run on an NVIDIA L40 GPU (46 GB VRAM). Additional implementation details, including network architecture and hyperparameters, are provided in the Supplementary Information.

To obtain the optimal reported reconstructions, TV regularization and early stopping were employed to mitigate overfitting and stabilize training. In the absence of a universally accepted stopping criterion for DIP reconstruction, an empirical strategy was adopted. Intermediate reconstructions were stored throughout optimization, and the optimal reconstruction was selected within the convergence plateau of the loss function based on the evaluation of representative cross-sections across iterations. This corresponded to the regime in which structural features stabilized and remained consistent across elemental channels.

## De-normalization and quantification:

Before reconstruction, the acquired sinograms were normalized independently by channel, resulting in reconstructed volumes expressed on a relative intensity scale (0, 1). Because quantitative EDX analysis requires preservation of physically meaningful X-ray count scaling, a de-normalization procedure was applied after reconstruction.

For each elemental channel, the DIPm-TV reconstruction was forward projected using the Radon transform with fine angular sampling (0.2°) and subsequently rescaled using the maximum experimental intensity of the corresponding sinogram. This procedure restores consistency with the experimental count scale while preserving the structural prior imposed by DIPm-TV.

To ensure compatibility with conventional quantification workflows, a final SIRT reconstruction was performed using the de-normalized projections. Owing to its linearity, SIRT preserves the relative intensity relationships introduced by DIPm-TV while maintaining physical consistency with the measured data. Semi-quantitative elemental concentrations were subsequently estimated using the CL formalism[30,34] implemented in HyperSpy[45].

# Data Availability:

The data supporting the findings of this study are available from the corresponding author upon reasonable request.

# Code availability:

The code used to generate the results in this study is available at: https://github.com/CEA-MetroCarac/DL_etomo

# Acknowledgments

Part of this work was carried out on the Platform for Nanocharacterisation (PFNC), supported by the “Recherche Technologique de Base” and "France 2030 - ANR-22-PEEL-0014" programs of the French National Research Agency (ANR). The authors also acknowledge the French National Research Agency through the PEPR DIADEM.

# Supplementary Information:

## Electron fluence calculation:

The electron fluence per pixel in scanning transmission electron microscope (STEM) is given by the following equation[1]:

$$Fluence\left[\frac{e^-}{\text{Å}^2}\right] = \frac{I[A] * t_{dwell}[s]}{e[C] \times d^2[\text{Å}^2]}, (1)$$

where $e$ is the electronic charge of an electron ($1.602 \times 10^{-19}C$), $I$ is the probe current, $t_{dwell}$ is the pixel dwell time and $d$ is the pixel size.

## Signal to noise ratio (SNR) estimation:

The signal-to-noise ratio (SNR) was estimated independently for each elemental map and projection angle. For each projection image, two regions were defined: a background region, consisting of lateral strips at the left and right edges of the image and verified to contain no sample signal, and a signal region defined as a centred rectangular region of interest (ROI), excluding both the background strips and an additional inner margin to avoid edge artefacts and transition zones. Within the signal ROI, an Otsu threshold was applied to separate sample pixels from residual background. The SNR was then computed as:

$$SNR_{dB} = 20\log_{10}\left(\frac{\mu_{signal} - \mu_{bg}}{\sigma_{bg}}\right), (2)$$

where $\mu_{signal}$ is the mean intensity of pixels above the Otsu threshold within the ROI, and $\mu_{bg}$ and $\sigma_{bg}$ are the mean and standard deviation of the lateral background strips, respectively.

| **Element** | **Virgin** | | **SET** | |
|---|---|---|---|---|
| | **Mean SNR (dB)** | **STD (dB)** | **Mean SNR (dB)** | **STD (dB)** |
| **Ge** | 29.5 | 2.2 | 27.9 | 0.4 |
| **Te** | 19.7 | 1.3 | 17.8 | 0.5 |
| **Sb** | 15.6 | 1.3 | 13.9 | 0.4 |

Table S1. Mean and standard deviation of the SNR (dB) for each elemental channel (Ge, Te, Sb) in the virgin and SET devices.

## Deep Image Prior with multi-channel formulation (DIPm-TV):

### Inverse problem and classical reconstruction approaches:

In electron tomography, the reconstruction problem is commonly formulated as a linear inverse problem:

$$y = Px + n \ (3)$$

where $y$ denotes the acquired projections, $x$ denotes the unknown 3D object to reconstruct, $P$ is the projection operator (discretized 3D Radon transform), and $n$ accounts for measurement noise. A widely used reconstruction algorithm is the Simultaneous Iterative Reconstruction Technique (SIRT), which minimizes a least-squares data fidelity term:

$$\hat{x} = argmin_x \frac{1}{2}\left|\left|Px - y\right|\right|_2^2 \quad (4)$$

When applied to limited-angle or sparse-view acquisitions, SIRT produces degraded reconstructions due to the ill-posed nature of the inverse problem. In order to mitigate these limitations, a common alternative is Compress Sensing (CS)-based approaches, which incorporate a regularization term promoting the sparsity of the object in a chosen transformation domain L (i.e., gradient, wavelets, etc.)[2–4]. This leads to the following regularized minimization problem:

$$\hat{x} = argmin_x \left\{\frac{1}{2}\left|\left|Px - y\right|\right|_2^2 + \lambda \cdot \left|\left|Lx\right|\right|_1\right\} \quad (5)$$

where the first term corresponds to the data-fidelity term, the second term corresponds to the regularization term, and $\lambda$ is a parameter that controls the trade-off between them.

In this work, we target piecewise-constant reconstructions, which are well promoted by total variation (TV) regularization, defined as the l1-norm of the gradient: $TV(x) = \left|\left|\nabla x\right|\right|_1$[4]. The resulting CS-TV reconstruction problem becomes:

$$\hat{x} = argmin_x \left\{\frac{1}{2}\left|\left|Px - y\right|\right|_2^2 + \lambda \cdot \left|\left|\nabla x\right|\right|_1\right\} \quad (6)$$

While CS-TV improves reconstruction quality under moderate limited-angle, it remains insufficient under the severe missing wedge (MW) conditions considered in this work.

### Deep image prior with TV regularization formulation:

To address this limitation, we adopt a DIP-TV strategy[5–7], where the unknown 3D volume is parameterized by a convolutional neural network (CNN):

$$x = F_\theta(z) \quad (7)$$

where $F_\theta\,(z)$ is a randomly initialized 3D CNN, and $z$ is a fixed random input tensor. Unlike supervised learning approaches, DIP-TV reconstructs the volume by directly optimizing the network parameters to fit the measured projections.

Under this parametrization, the reconstruction is obtained by optimizing the network parameters $\theta$:

$$\hat{\theta} = argmin_\theta \left\{\frac{1}{2}\left|\left|PF_\theta(z) - y\right|\right|_2^2 + \lambda \cdot \left|\left|\nabla F_\theta(z)\right|\right|_1\right\}, \quad (8)$$

where the first term corresponds to an $\ell_2$ data fidelity term (equivalent to a mean squared error up to a constant factor), and the second term enforces TV regularization. The reconstructed object is given by

$$\hat{x} = F_{\hat{\theta}}(z). \quad (9)$$

This formulation leverages the implicit bias of convolutional architectures to favour structured, natural solutions, thereby mitigating artefacts associated with ill-posed inverse problems[8].

## Multi-channel DIP-TV formulation (DIPm-TV):

We extend this DIP-TV strategy to a multi-channel formulation (DIPm-TV), enabling the joint reconstruction of multiple spectroscopic signals. Let $y^c$ and $F_\theta^c(z)$ denote the projections and reconstructed volume for channel $c \in \{1, \cdots, N\}$, where $N$ is the total number of channels.

A single 3D U-Net outputs all channels simultaneously:

$$F_\theta(z) = \left[F_\theta^1(z), \cdots, F_\theta^N(z)\right]. \ (10)$$

The optimization problem becomes:

$$\hat{\theta} = \underset{\theta}{\operatorname{argmin}} \sum_{c=1}^{N} \left\{\frac{1}{2}\left|\left|PF_\theta^c(z) - y^c\right|\right|_2^2 + \lambda \cdot \left|\left|\nabla F_\theta^c(z)\right|\right|_1\right\}, \ (11)$$

and the reconstructed multi-channel volume is expressed as:

$$\hat{x} = \left[F_{\hat{\theta}}^1(z), \cdots, F_{\hat{\theta}}^N(z)\right]. \ (12)$$

Note that for $N = 1$, (11) reduces to the standard single-channel DIP-TV formulation[5,6].

This joint reconstruction exploits shared spatial correlations across channels, improving interface definition and reducing noise. In particular, low-SNR channels benefit from structural information provided by higher-SNR channels, leading to enhanced consistency across elemental maps.

## DIP-TV on a synthetic phantom:

The structure consists of a square-based cuboid “matrix” region with sharp edges, inside which two inclusions are embedded: a sphere and a flattened ellipsoid elongated along one in-plane axis (Figure S1). The matrix signal was defined as a homogeneous high-intensity region with voids at the inclusion sites, while the remaining signals represent the inclusions with constant and distinct intensities.

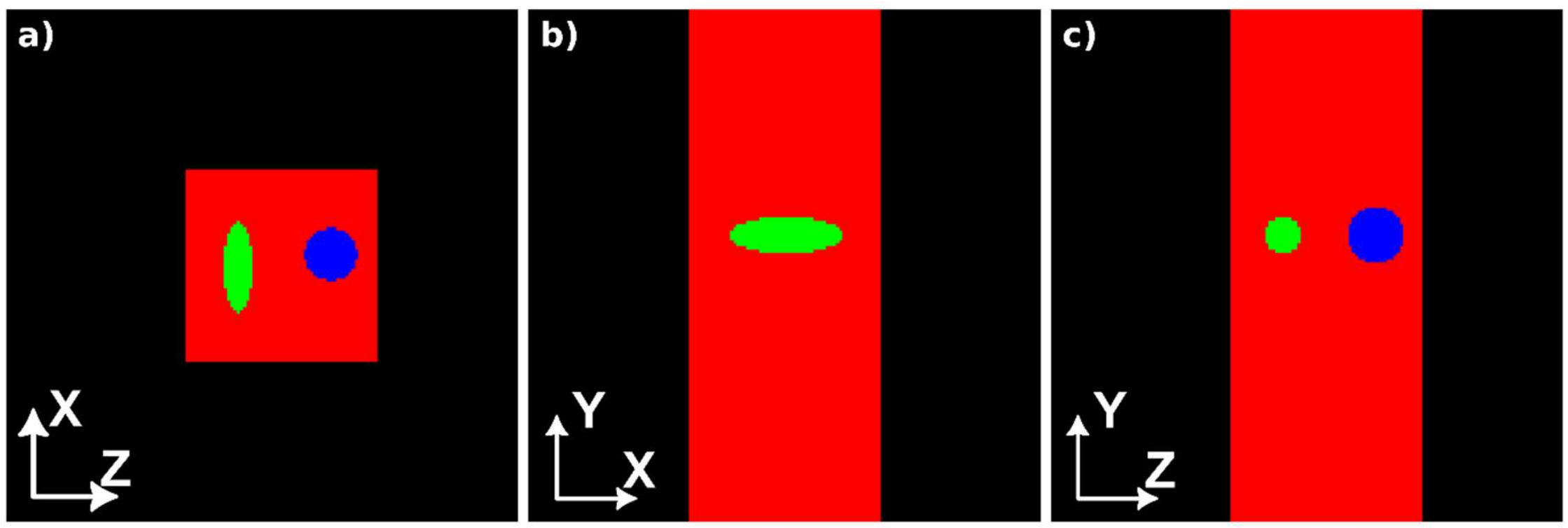

Figure S1. Cross-sections views of the simulated phantom.

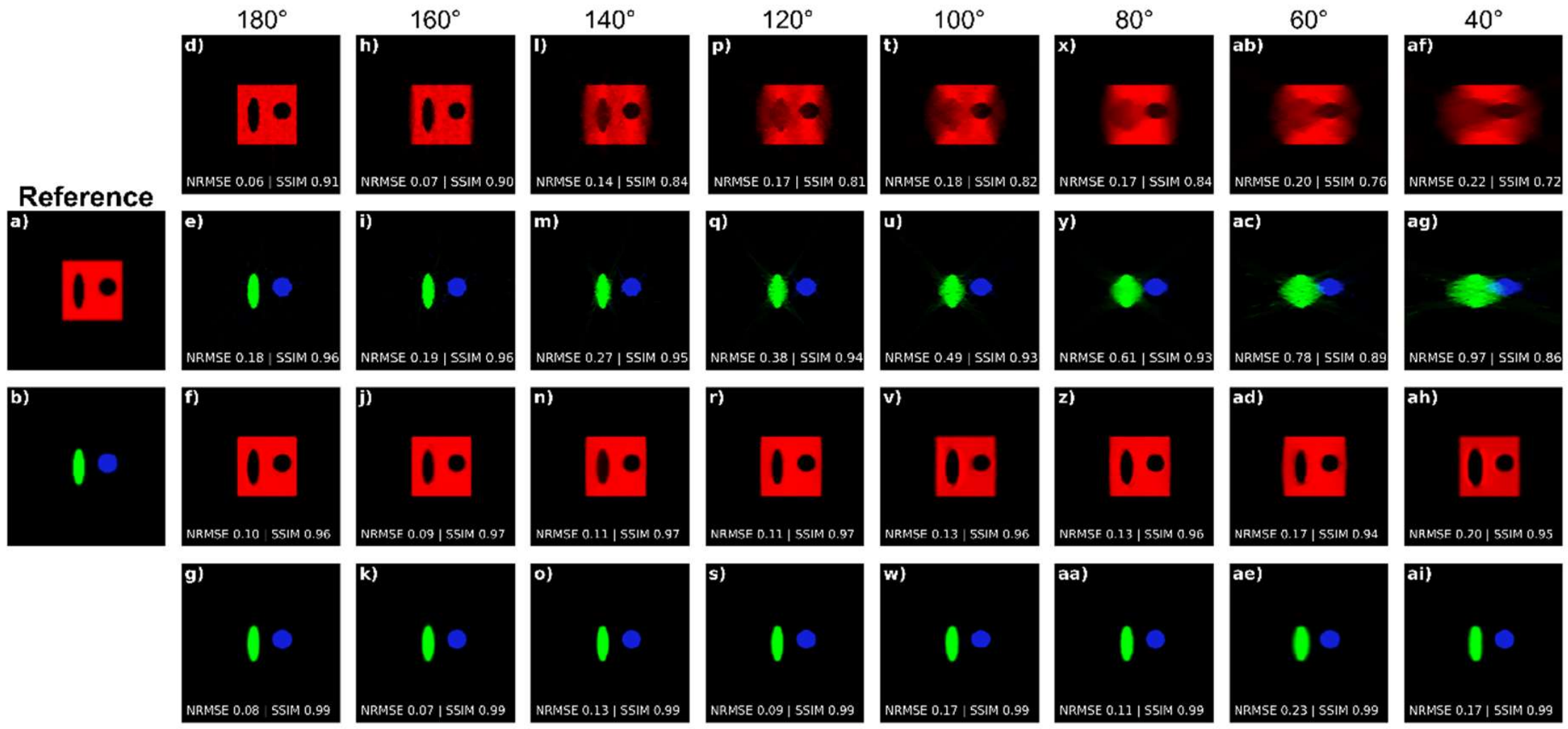

Figure S2. Influence of limited tilt range on simulated phantom reconstructions. (a-b) Reference slices of the three-channel phantom. (d-ai) Representative cross-section slices for all the tilt ranges evaluated: upper rows correspond to CS-TV reconstructions, and lower rows to DIPm-TV reconstructions. The MW is displayed horizontally

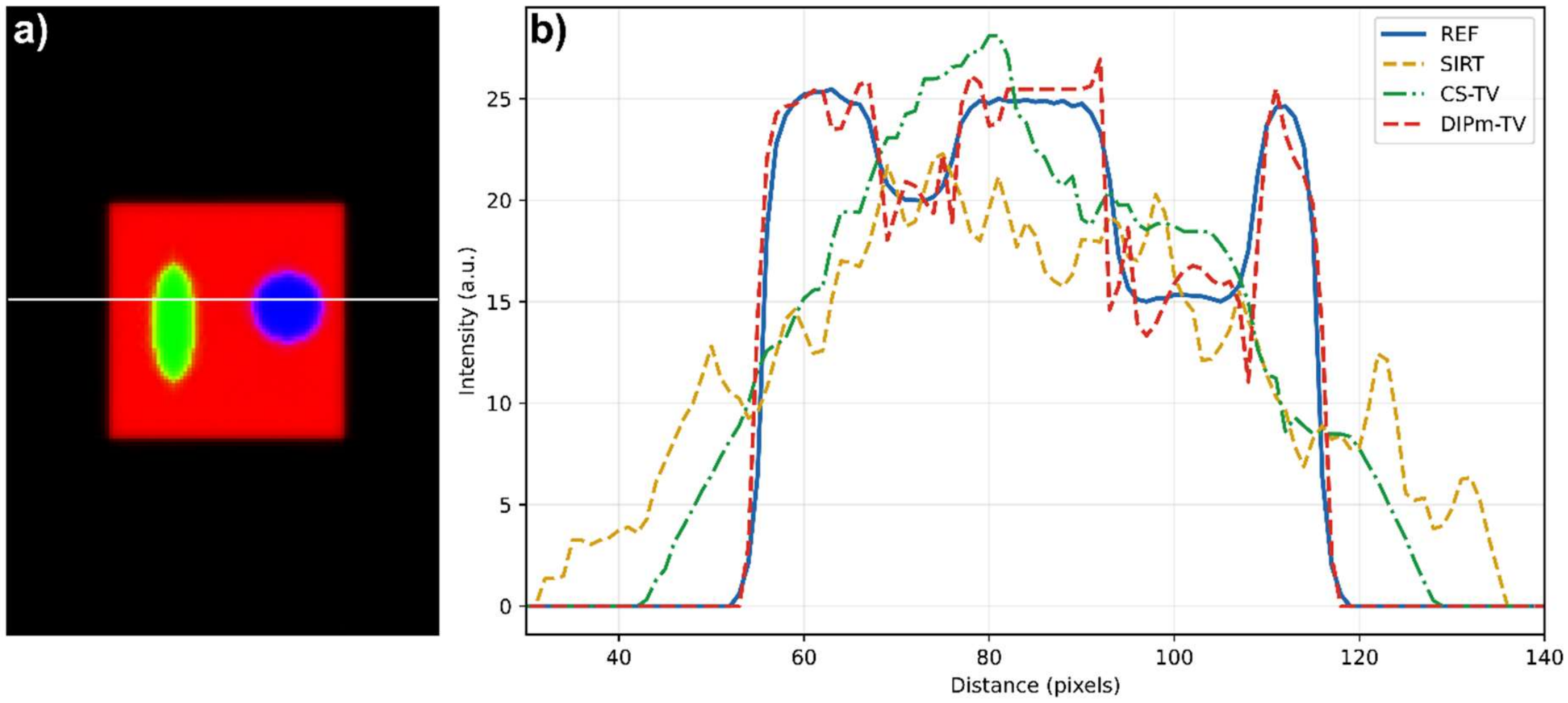

Figure S3. (a) Central slice of the simulated phantom, with the horizontal line indicating the position of the extracted profile. (b) Comparison of line profiles extracted along the same cross-section shown in (a) for the reference phantom (REF, blue), SIRT (yellow), CS-TV (green), and DIPm-TV reconstruction (red). The data correspond to a limited tilt range scenario of [-40°:5°:+40°] with added noise.

# Power spectral density (PSD)-based resolution estimation:

To quantify directional resolution in the reconstructed volumes, we estimated a spatial-frequency cut-off derived from the power spectral density (PSD)[9] along each axis. One-dimensional PSD profiles were computed along the x, y and z directions. For each axis, the 1D PSD was obtained by Fourier transforming the reconstructed volume and extracting line profiles through the origin in the corresponding frequency direction. To reduce variance, these profiles were averaged over neighbouring directions. Each 1D PSD was then fitted in log space using a Lorentzian-like roll-off model, together with a constant offset accounting for an approximately white noise floor:

$$PSD(k) = \frac{A}{[1 + (2\pi k\xi)^2]^p} + C \quad (8)$$

where $A$, $\xi$ and $p$ describe the signal decay and $C$ accounts for the high-frequency noise floor.

The noise-dominated tail was modelled by fitting a log-linear function, $log_{10}N(k) = a + b \cdot k$, to the final portion of the PSD tail; yielding $N(k) = 10^{a+b\cdot k}$. The cut-off frequency $k_{cut}$ was defined as the intersection between the fitted PSD model and the estimated noise floor, i.e., $PSD(k_{cut}) = N(k_{cut})$. This provides an operational resolution limit beyond which the reconstruction is dominated by noise and high-frequency details cannot be reliably recovered [9]. Finally, the effective spatial resolution was computed as $r \approx 1/k_{cut}$ (in nm, with $k$ expressed in $nm^{-1}$).

| Method/Signal | | Resolution XY (nm) | Resolution Z (nm) | Difference (nm) | Ratio |
|---|---|---|---|---|---|
| **SIRT** | **Ge** | 5.8 | 7.8 | 2.0 | 1.35 |
| | **Te** | 5.6 | 7.2 | 1.6 | 1.30 |
| | **Sb** | 5.8 | 7.8 | 2.0 | 1.34 |
| **CS-TV** | **Ge** | 5.3 | 7.0 | 1.7 | 1.32 |
| | **Te** | 5.6 | 7.2 | 1.6 | 1.30 |
| | **Sb** | 5.7 | 7.4 | 1.7 | 1.29 |
| **DIPm-TV** | **Ge** | 5.2 | 5.7 | 0.6 | 1.08 |
| | **Te** | 5.3 | 5.7 | 0.4 | 1.06 |
| | **Sb** | 5.5 | 5.9 | 0.4 | 1.07 |

Table S2. Power spectral density-based spatial resolution for each element. The lateral (XY) resolution is reported as the mean of the cut-off resolutions obtained along $k_X$ and $k_Y$, while the axial (Z) resolution corresponds to the cutoff along $k_Z$ (MW direction). The difference denotes $\Delta = Res_Z - Res_{XY}$ and the ratio denotes the $Res_Z/Res_{XY}$.°

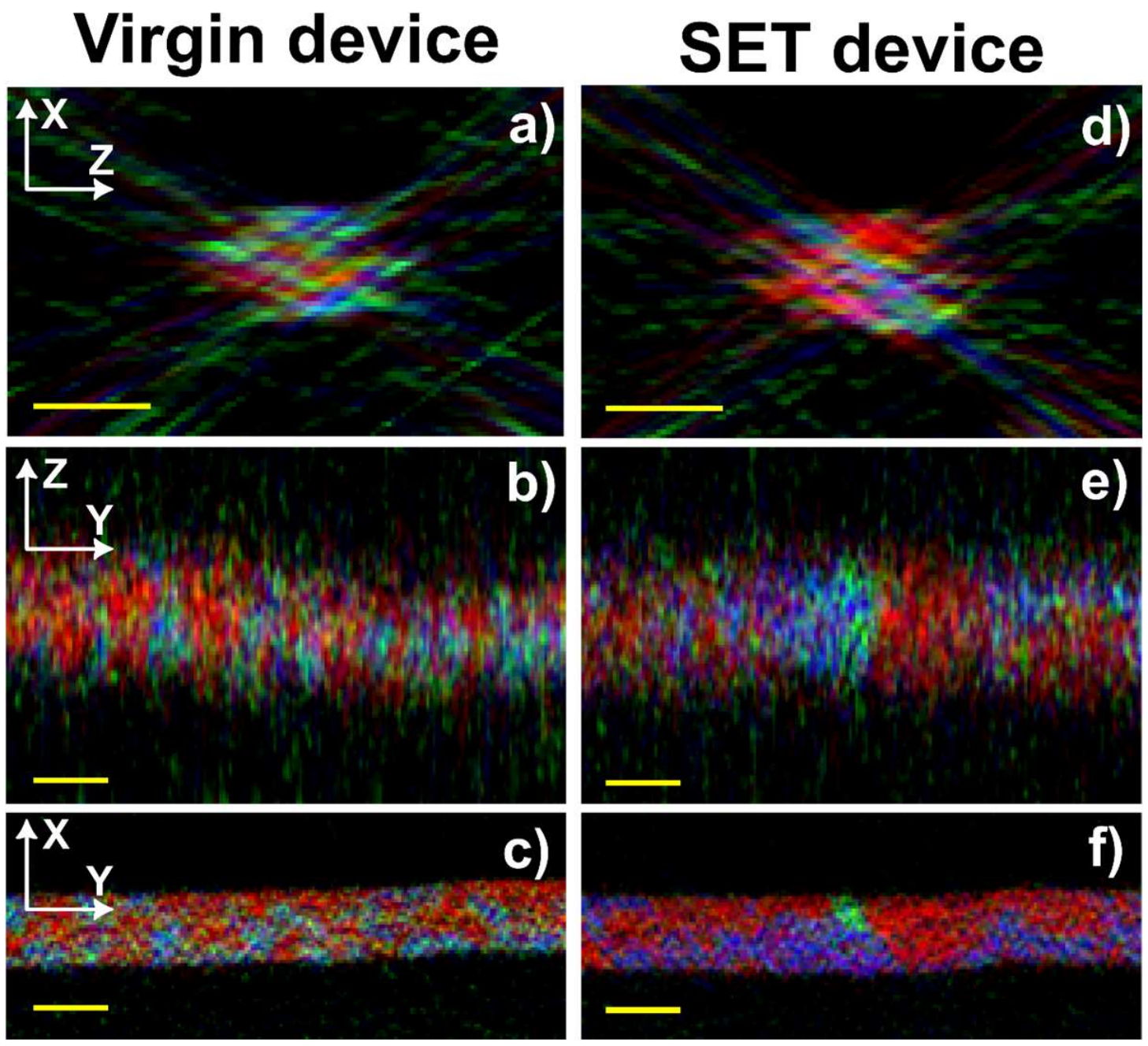

Figure S4. Representative XZ, YZ, and XY orthoslices through the SIRT reconstructions of the virgin (left) and SET (right) devices. Scale bars: 50 nm. The MW is in the Z direction.

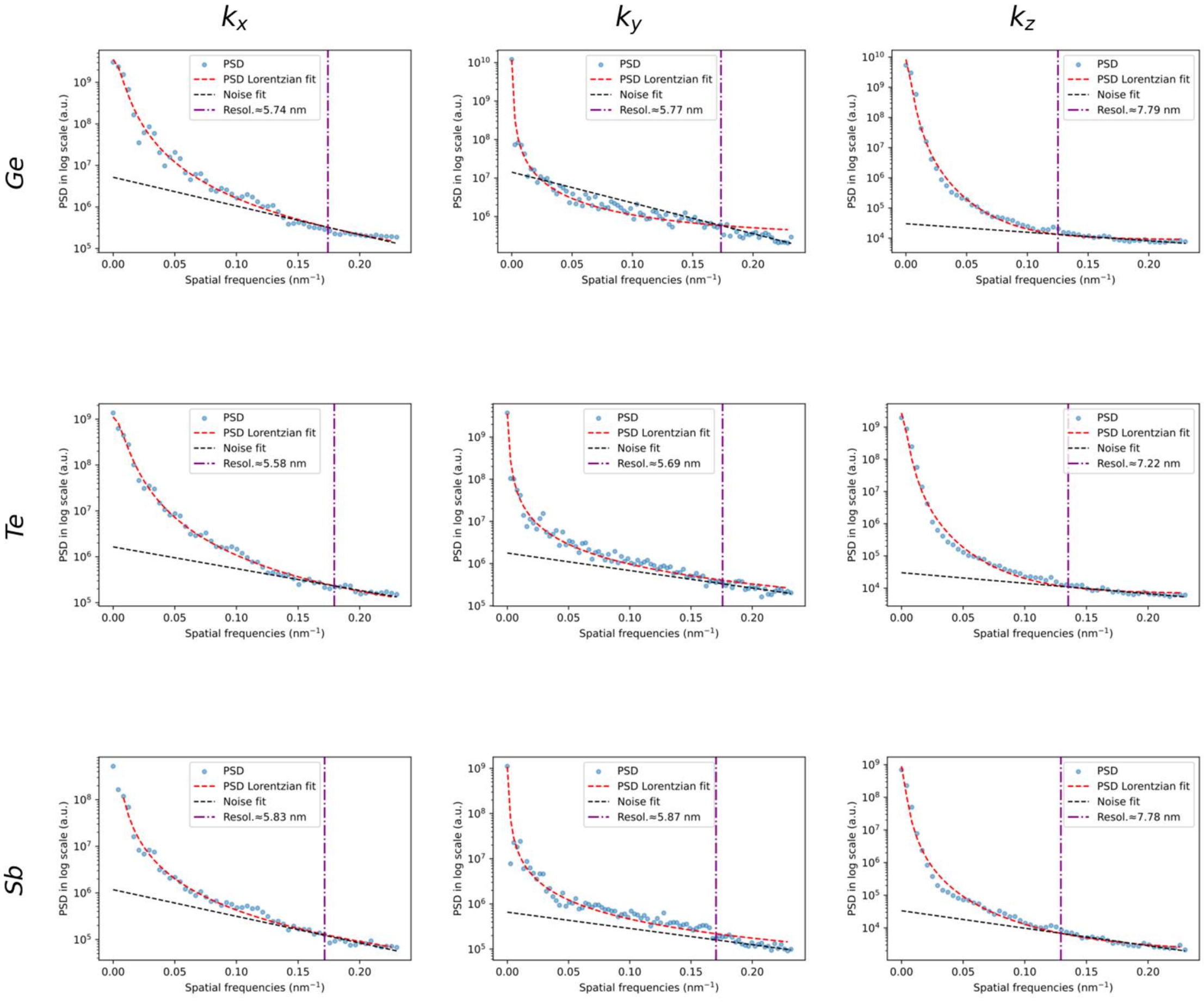

Figure S5. PSD-derived directional spatial-frequency cut-offs from 1D PSD profiles computed along $k_x$, $k_y$ and $k_z$ to assess directional resolution for the SIRT reconstructions of the virgin PCM device. Each panel shows the PSD profile with a Lorentzian fit (red) and a power law fit to the noise tail (black). The intersection defines the cut-off frequency, from which the corresponding spatial resolution is calculated.

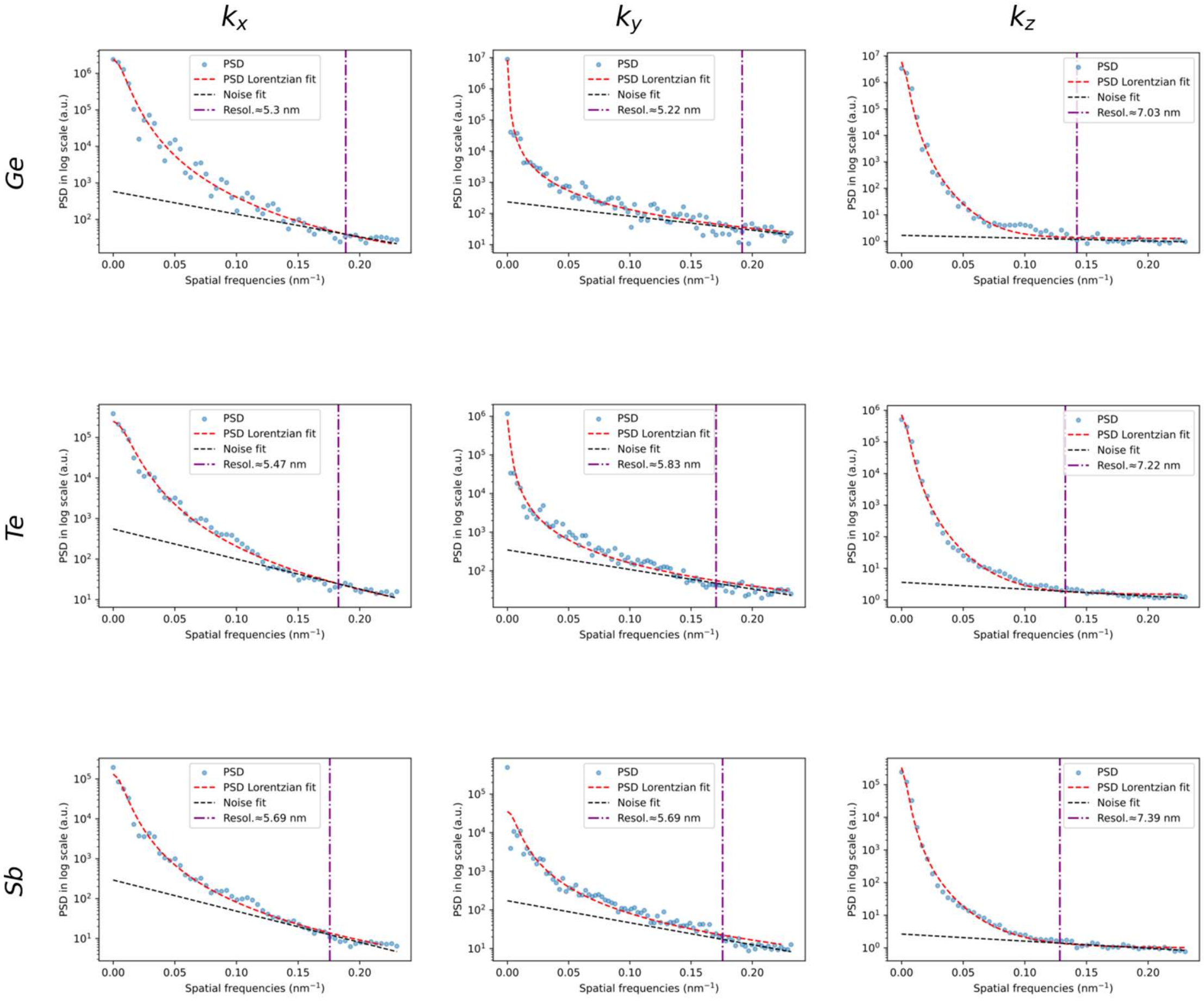

Figure S6. PSD-derived directional spatial-frequency cut-offs from 1D PSD profiles computed along $k_x$, $k_y$ and $k_z$ to assess directional resolution for the CS-TV reconstructions of the virgin PCM device. Each panel shows the PSD profile with a Lorentzian fit (red) and a power law fit to the noise tail (black). The intersection defines the cut-off frequency, from which the corresponding spatial resolution is calculated.

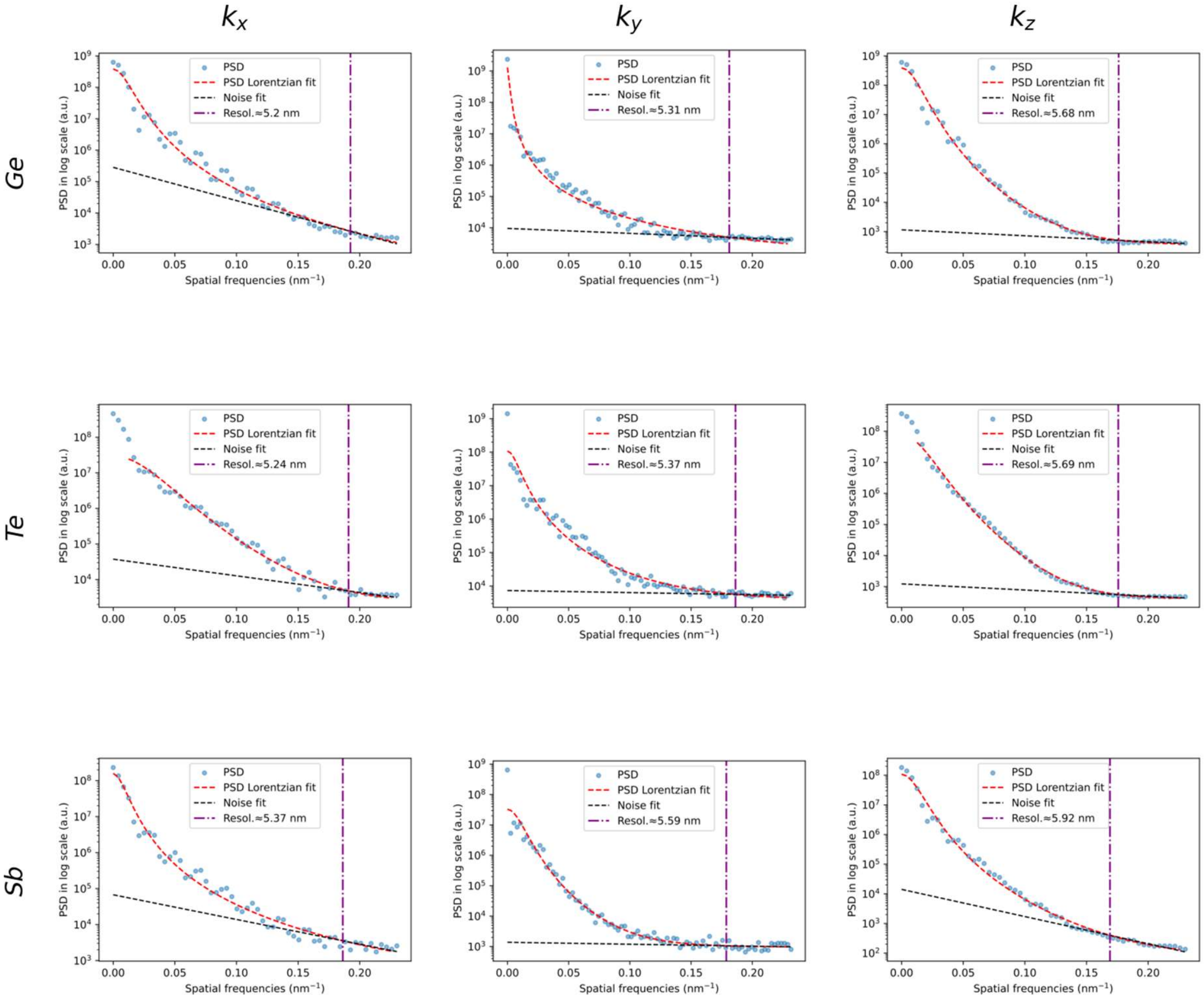

Figure S7. PSD-derived directional spatial-frequency cut-offs from 1D PSD profiles computed along $k_x$, $k_y$ and $k_z$ to assess directional resolution for the DIPm-TV reconstructions of the virgin PCM device. Each panel shows the PSD profile with a Lorentzian fit (red) and a power law fit to the noise tail (black). The intersection defines the cut-off frequency, from which the corresponding spatial resolution is calculated.

## Quantitative compositional analysis:

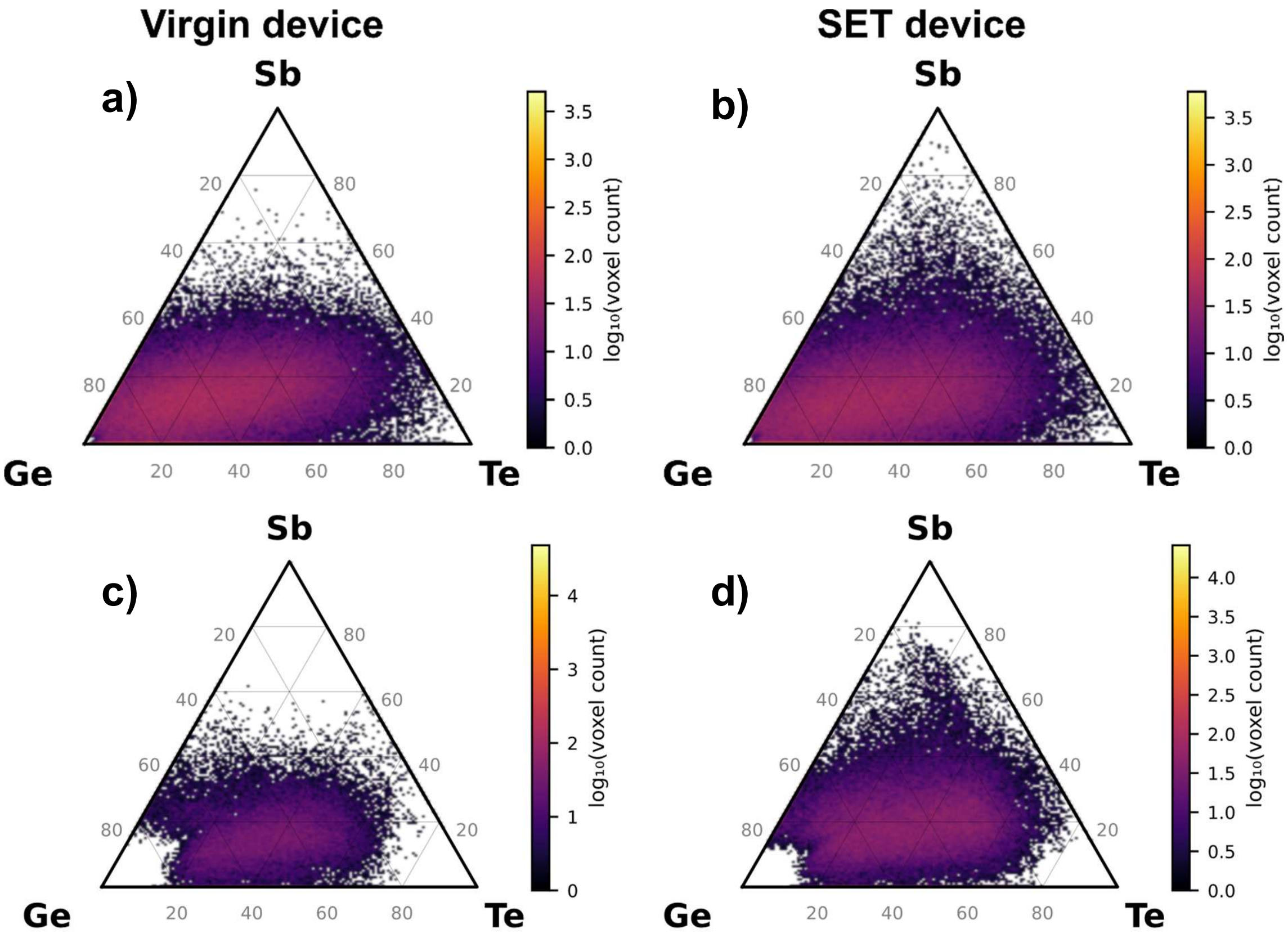

Figure S8. Ternary composition diagrams of the reconstructed volumes for the virgin and SET PCM devices using SIRT ((a, b), respectively) and CS-TV ((c, d), respectively). The voxel composition distributions are represented in the Ge-Sb-Te compositional space, where the colour indicates the logarithm of voxel counts.

| Sample | Cluster | Ge (at. %) | Sb (at. %) | Te (at. %) | Sb/Te |
|---|---|---|---|---|---|
| **Virgin** | **C1** | 78.6 | 10.2 | 11.1 | 0.92 |
| | **C2** | 31.1 | 19.3 | 49.6 | 0.39 |
| **SET** | **C1** | 78.1 | 11.3 | 10.6 | 1.06 |
| | **C2** | 36.4 | 18.8 | 44.9 | 0.42 |
| | **C3** | 19.3 | 52.9 | 27.9 | 1.89 |

Table S4. Elemental quantification for the densest region per cluster identified in the ternary plot of Figure 4.

## DIPm-TV training and implementation details:

The main DIP-TV training hyperparameters are summarized in Table S5, including the input noise amplitude, input depth, learning rate, noise regularization amplitude, number of iterations, and the TV regularization weight ($\lambda_{TV}$). Hyperparameters were explored over the ranges indicated below, and final values were selected based on reconstruction quality and stability.

| DIP training parameters | |
|---|---|
| **Initial input noise amplitude** | 0.1 |
| **Input depth** | 32 |
| **AdamW learning rate** | $10^{-3} - 10^{-5}$ |
| **Noise regularization value** | $0.01 - 0.1$ |
| **Number total of iterations** | 2000 |
| $\boldsymbol{\lambda_{TV}}$ | $10^{-7} - 10^{-11}$ |

Table S5. DIP-TV training hyperparameters explored in this work. Final values were selected within these ranges depending on the dataset and noise level.

The neural network $F_\theta$ was implemented in PyTorch as a 3D U-Net-like architecture. Downsampling was performed using strided 3D convolutions, while upsampling relied on trilinear interpolation followed by convolution to reduce checkerboard artefacts. LeakyReLU activations and batch normalization were used throughout the network, while reflection padding was applied to reduce boundary artefacts. The encoder-decoder filter configuration consisted of [64,128,256,256], with skip filters [16,32,64,64]. To discourage degenerate zero-valued solutions, a "force non-zeros" strategy was adopted by squaring the network output.

The network input consisted of fixed random noise sampled from a uniform distribution, with dimensions matching the reconstruction volume and an input depth of 32 channels. To improve robustness and reduce overfitting, noise regularization was introduced by perturbing the fixed input at each iteration with additive zero-mean Gaussian noise of controlled amplitude[5].